# Hydrogen-bonded aggregates in the mixtures of piperidine with water: thermodynamic, SANS and theoretical studies

Wojciech Marczak[1,2*], Jakub T. Hołaj-Krzak[3], Piotr Lodowski[3], László Almásy[4], Giulia C. Fadda[5]

[1]Institute of Occupational Medicine and Environmental Health, Kościelna 13, 41-200 Sosnowiec, Poland

[2]Department of Physical Chemistry, Kazan Federal University, Kremlevskaya str. 18, 420008 Kazan, Russia

[3]Institute of Chemistry, University of Silesia, Szkolna 9, 40-006 Katowice, Poland

[4] Wigner Research Centre for Physics, POB 49, Budapest – 1525, Hungary

[5] Laboratoire Léon Brillouin CEA-CNRS, 91191 Gif sur Yvette Cedex, France

**Abstract**

Structures resembling semiclathrates probably arise in liquid aqueous solutions of piperidine at the amine mole fraction below 0.03. With the increasing concentration, the structures gradually decay, but the 1:1 complexes of piperidine with water remain linked one to another through the O–H...O bonds between the hydration water molecules. A periodic order of the bicontinuous microemulsion type occurs in the range of the mole fractions from 0.08 to 0.5. In the piperidine-rich mixtures, the 1:1 complexes are dispersed uniformly in the amine. Relatively low stabilization energy of these complexes probably causes that piperidine is totally miscible with water.

[*] Corresponding author

E-mail address: w.marczak@imp.sosnowiec.pl (W. Marczak)

# 1. Introduction

Unambiguous determination of the liquid structure, similar to that achieved by the X-ray crystallographic methods for solids, is impossible at the present state of knowledge. However, intuitive molecular models of liquids often refer to ordered solid phases. Recently, piperidine and *N*-methylpiperidine were reported as promoters of the methane clathrate hydrate formation. Piperidine with methane forms sII clathrate hydrate in aqueous solutions at elevated pressures [1]. Slightly larger molecules of methylpiperidines cause the formation of sH clathrates in similar conditions [2]. Permanent interest in different clathrate hydrates is justified by their potential application as the gas-storage media [1, 3-5]. Although piperidine does not form solid clathrate hydrate without a help gas [1], it seems probable that water polyhedra similar to those in the solid clathrate hydrates may occur in its liquid aqueous solutions. Obviously, the clathrate-like structures in liquids would be dynamic molecular aggregates rather than the stable ones in the solid state [6-8].

Our previous studies pointed to different modes of hydration and association in the liquid system *N*-methylpiperidine – water [9]. At low amine concentrations, semiclathrate-like structures probably arise around the *N*-methylpiperidine molecules. Along with increased amine concentration, the clathrate polyhedra gradually decay, while the amine molecules remained O–H...N bonded to water. The RN...H–OH hydrates aggregate due to the O–H...O bonds between the hydrate water molecules. Thus, the hydrophilic interactions prevail in the amine-rich mixtures. The aggregation of the RN...H–OH hydrates is probably the reason of the phase splitting of the system above the lower critical solution temperature of 316.7 K at the amine mole fraction $x_1 = 0.06 \pm 0.01$ [9] (LCST 315 K at $x_1 = 0.07$ [10]). Contrary to *N*-methylpiperidine, piperidine is totally miscible with water at atmospheric pressure in the temperature range 273.15 – 368.15 K [11].

In the present paper, we report results of the small-angle neutron scattering (SANS) and thermodynamic studies of binary liquid system piperidine – water, supplemented by the H-bond energies and geometries of the molecular complexes calculated theoretically. We compared the present results with those obtained earlier for *N*-methylpiperidine + water [9]. Considering the molecular interactions, an important difference between *N*-methylpiperidine and piperidine is the ability of the latter to self-association due to the N–H...N bonds. The goal was to find how is that manifested in the studied thermodynamic and structural properties of the two aqueous systems. Although the density of aqueous solutions of piperidine and the speeds of sound have been reported already [12-14], we decided to repeat the acoustic and volumetric measurements for the sake of consistency with the data for the *N*-methylpiperidine system.

## 2. Experimental

*2.1. Chemicals*

Piperidine (Aldrich, manufacturer's assay 99 %) was distilled using a rectifying column filled with glass rings. Boiling temperature of the collected fraction was equal to the value recommended by NIST, (379.2 ± 0.5) K [15], within the measurement uncertainty limits of 1 K. During distillation, vapors were in contact with the drying agent $Na_2O$. The distillate was stored over molecular sieves 4 Å. Chemicals from two batches were purified in the same manner.

Contaminants: 2-methylpiperidine and 1-ethylpiperidine were detected in the chemical by the GCMS analysis. However, the measured densities remained in good agreement with those reported in the literature [13, 16]. Differences between the experimental densities and the literature values in the temperature range (273.15 – 323.14) K were from (0.13 to 0.17) kg m$^{-3}$ [13] and from (0.36 to 0.04) kg m$^{-3}$ [16] for the first distillate, while from (–0.29 to –0.23) kg m$^{-3}$ [13] and from (–0.06 to –0.36) kg m$^{-3}$ [16] for the second distillate. Other reported densities of piperidine were by ca. 2 kg m$^{-3}$ higher [12]. Dry piperidine easily absorbs atmospheric moisture, that causes an increase of the liquid density. For that reason, we used the distillate with lower density in the thermodynamic experiments.

Water ($H_2O$) was double-distilled and de-gassed by boiling just before preparation of the solutions. Its electrolytic conductivity was $1.5 \times 10^{-4}$ S m$^{-1}$.

Heavy water (Sigma, 99.9 % $D_2O$) was used without any pre-treatment.

The solutions were prepared by mass, using analytical balances with accuracy of $\pm 0.6 \times 10^{-3}$ g. The uncertainty in mole fractions did not exceed $5 \times 10^{-5}$. Before the speed of sound and density measurements, the samples were de-gassed in an ultrasonic cleaner.

*2.2. Apparatus*

Small-angle neutron scattering experiments were performed on the PACE instrument in the Laboratoire Léon Brillouin. The sample-detector distance and the incoming wavelength were 0.9 m and 0.45 nm that provided *q* range from 0.5 to 5 nm$^{-1}$. The samples were placed in 2-mm thick quartz cuvettes. The raw data were processed according to the procedure suggested by Brûlet et al. [17]

Density was measured with a vibrating-tube densimeter Anton Paar DMA 5000, with precision of $5 \times 10^{-3}$ kg m$^{-3}$ and accuracy by one order of magnitude lower. Experimental details were given earlier [18].

The group speed of ultrasound at $f = 4$ MHz was measured using a sing-around apparatus with precision of $\pm 0.05$ m s$^{-1}$ and the accuracy by one order of magnitude worse. Re-distilled water was used as a standard liquid for the calibration [19] at 11 temperatures differing by about 2 K in the range 293 – 313 K. Temperature in the ultrasonic cell was measured with a Pt100 resistance thermometer Elmetron PT-401, calibrated at the freezing and boiling points of water. Details of similar ultrasonic measurements were reported earlier [20].

## 3. Experimental results

Neutron scattering was measured at temperature 298.15 K in seven solutions of piperidine in D$_2$O. The mole fraction of piperidine $x_1$ ranged from 0.041 to 0.589. All the mixtures exhibit coherent small-angle scattering signal, indicating that the structure is heterogeneous on the nanometer length scales.

A good fit to all scattering data was obtained using the Teubner-Strey scattering function, which describes a micro-phase separated system with two characteristic lengths [21,22]:

$$I = \frac{1}{a_2 + c_1 q^2 + c_2 q^4} + I_{bg}. \tag{1}$$

The weighted least squares fitting was performed using the SASFit software [23], including corrections for the instrumental smearing. From the fitted coefficients $a_2$, $c_1$, and $c_2$, the quasi-periodic repeat distance $D$ and the decay length of the correlations $\xi$ were calculated:

$$D = 2\pi \left[ \frac{(a_2/c_2)^{1/2}}{2} - \frac{c_1}{4c_2} \right]^{-1/2}, \tag{2}$$

$$\xi = \left[ \frac{(a_2/c_2)^{1/2}}{2} + \frac{c_1}{4c_2} \right]^{-1/2}. \tag{3}$$

The scattering curves are plotted in Figure 1 and the characteristic lengths $D$ and $\xi$ are reported in Table 1.

The mixture of the lowest concentration, $x_1 = 0.041$, shows enhanced forward scattering which can be modeled also by the conventional Ornstein-Zernike type of scattering function, *i.e* by Eq. (1) with $c_2 = 0$, which is basically a Lorentzian peak function centered at $q = 0$. Deviation from this behavior in a form of an interference peak is pronounced for all the other mixtures. The peak gradually shifts towards larger $q$ values with increasing concentration.

Densities ($\rho$) of binary mixtures of piperidine with $H_2O$ were measured in the temperature range 288 – 313 K, while those of the two samples of pure piperidine at temperatures 273.15 – 323.15 K. Speed of sound ($u$) was measured in the temperature range 293 – 313 K. Raw results and fitted equations of the temperature dependencies are reported in the supplementary materials. Those equations were applied in all the subsequent thermodynamic calculations,

Isentropic compressibilities, $\kappa_S$, were calculated from the Laplace's formula:

$$\kappa_S = \rho^{-1} u^{-2}, \tag{4}$$

and they are plotted in Figure 2.

Thermodynamic excesses of molar functions: volume $V$, isobaric expansion $E_p = (\partial V/\partial T)_p$, and isentropic compression $K_S = \kappa_S V$, were calculated from the thermodynamically exact formulas in the same way as that reported earlier [9]. Indispensable in calculations isobaric heat capacities of piperidine and water were taken from the literature [24, 25] and the values of 85.1475 g mol$^{-1}$ and 18.01524 g mol$^{-1}$ of the molar masses were used.

The following empirical polynomials, similar to those suggested for the system *N*-methylpiperidine + water [9], were fitted to the excess volumes and compressions:

$$\frac{Y^E}{x_1 x_2} = \sum_{i=0}^{4}\sum_{j=0}^{1} a_{ij} c^i \vartheta^j, \tag{5}$$

where $Y = V \times 10^6/(\text{m}^3\ \text{mol}^{-1})$ or $Y = K_S \times 10^{15}/(\text{m}^3\ \text{mol}^{-1}\ \text{Pa}^{-1})$, $c = x_1/(x_1 + x_2/5)$ and $\vartheta = T/\text{K} - 298.15$. The regression coefficients $a_{ij}$ are attached in the supplementary materials. They were calculated by the least squares method, applying a stepwise rejection procedure. From the initial sets of 20 coefficients for the $V^E$ (maximum $i = 4$ and $j = 3$) and 24 for the $K_S^E$ function (maximum $i = 5$ and $j = 3$), only 9 and 7 coefficients, respectively, proved to be statistically significant. The $V^E(x_1)$ and $K_S^E(x_1)$ isotherms are plotted in Figure 3. The former agree with those reported by Afzal et al. [13].

The excess isobaric expansion was calculated directly from the molar expansions and as the temperature derivative of the excess volume:

$$E_p^E = \left(\partial V^E/\partial T\right)_p \tag{6}$$

with the $V^E(T,x_1)$ function given by Eq. (5). The results are shown in Figure 4. It is evident that the calculations with the interpolating Eq. (5) leads to the function that only semi-quantitatively approximates the $E_p^E$ values in the studied range of temperature. That is due to the asymmetry of the $V^E(T)$ curve and because the excess volume depends on temperature rather weakly. Similar semi-quantitative result was obtained for the system with $N$-methylpiperidine [9].

The partial molar volumes and isentropic compressions of piperidine were calculated from the excess functions given by Eq. (5) according to the formula:

$$Y_1 = Y^{id} + Y^E + (1-x_1)(\partial Y^{id}/\partial x_1 + \partial Y^E/\partial x_1) \tag{7}$$

and from the following approximate relationship:

$$Y_1 \approx \overline{Y} + (1-\overline{x}_1)(\Delta Y/\Delta x_1), \tag{8}$$

where $\overline{Y}$ and $\overline{x}_1$ are the arithmetic means of the two consecutive experimental values of the molar function (compression $K_S$ or volume $V$) and of the mole fraction $x_1$, respectively. $\Delta Z$ and $\Delta x_1$ are the appropriate differences. The dependencies of partial functions on concentration are plotted in Figure 5.

## 4. Theoretical Calculations

Interaction energies in the hydrogen-bonded complex of piperidine with water and in the piperidine dimer were calculated theoretically from the density functional theory (DFT) and the second-order Møller-Plesset perturbation theory (MP2) using the Dunning's augmented correlation consistent polarized valence double-ξ (aug-cc-pVDZ) basis set [26, 27]. In the DFT calculations, the B3LYP exchange-correlation functional was employed [28, 29]. The geometries of the isolated monomers and the complexes were fully optimized at the DFT/B3LYP level of theory. Those geometries were adopted in the calculations of the structure energies in the MP2 method. The energies and geometries were also calculated using double hybrid B2PLYP method, which combine the exact HF exchange with a MP2-like correlation to a DFT [30]. Details of the calculations using Gaussian09 program package [31] were given in a previous work [32].

The association energies and the selected geometrical parameters around hydrogen bond are reported in Table 2, and the optimized structures of molecular complexes are presented in Figure 6. The MP2/aug-cc-pVDZ stabilization energy of the $C_5H_{10}NH \cdot H_2O$ complex is by 3.9 kJ mol$^{-1}$ lower than 25.8 kJ mol$^{-1}$ obtained at the MP2/6-31G** level of theory [33]. Similarly as for the system *N*-methylpiperidine – water [9], the MP2 energies are higher than the DFT/B3LYP ones. That is probably because of the underestimated dispersion energy in the DFT results [34, 35].

5. Discussion

The energies of water – piperidine interactions are considerably higher than those in the dimers of piperidine (Table 2) or water (9.5 kJ mol$^{-1}$ [34] and 12.17 kJ mol$^{-1}$ [36]; MP2 with BSSE and ZPE corrections). Thus, the cross-associates are formed when piperidine is diluted in water, while the concentration of the self-associated molecules decreases. As a result, the dilution of piperidine in water is strongly exothermic, with the limiting partial molar excess enthalpy of the amine equal to –25.94 ± 0.4 kJ mol$^{-1}$ at $T$ = 298.15 K [37]. For the same reason, the mean intermolecular distances are shorter in the real mixture than in the respective thermodynamically ideal one, that is manifested in the negative excess volume. The negative excess compression (Figure 3) results not only from the close-packing of the molecules, but also point to the mixture "structuredness" enhanced by the O–H...N bonds. Free proton in water molecule of the aggregate is easily accessible for an acceptor molecule. The electron pairs are also exposed (Fig. 6). Thus, the cross-associates tend to form larger aggregates due to the O–H...O bonds between the hydrate water molecules in the same way as that suggested for *N*-methylpiperidine [9] and pyridines [38].

Unfavorable entropy change caused by the aggregation is compensated by the increased energy of the O–H...O bonds in the vicinity of the O–H...N ones. Such non-additivity of the interactions energy is commonly known as the cooperativity of hydrogen bonds. The hydrogen-bonded aggregates dissociate with increasing temperature that causes positive excess of the thermal expansion $E_p^E$ at $x_1 > 0.015$ (Figure 4). The convexity of the excess molar expansion function at $x_1 < 0.015$, similar to that reported for the *N*-methylpiperidine – water system, probably results from the electrolytic dissociation [9].

Small-angle neutron scattering provided strong argument for the molecular aggregation in the solutions of piperidine in water. For the lowest concentration studied, $x_1$ = 0.041, two least squares fits of the Teubner-Strey model, Eq. (1), to the $I(q)$ scattering data are visually indistinguishable one from another: the first with adjustable $c_2$ coefficient, and the second with $c_2$ set to zero. The latter is the conventional Ornstein-Zernike scattering function. Such Ornstein-Zernike scattering is typical for binary solutions which are close to the boundary of phase separation [9, 38-42]. While piperidine is miscible with water in the whole concentration range independently of temperature [11], binary aqueous systems of its methyl derivatives exhibit the lower consolute points. The LCSTs are 315 K [10] – 317 K [9] for *N*-methylpiperidine and 339 K for 2-methylpyridine [10] systems. The former show the Ornstein-Zernike scattering at room temperatures below the LCST in the *N*-methylpiperidine mole fraction range up to 0.1 [9]. It is worth noting that the MP2 association energy of the 1:1 *N*-methylpiperidine-water complexes, 24.0 kJ mol$^{-1}$ [9], is higher than that for similar complexes of piperidine, 21.9 kJ mol$^{-1}$, reported in Table 2. Thus, the attraction between unlike molecules is stronger in the system with the miscibility gap, and the maximum negative molar excesses of volume and isentropic compression are bigger: $|V_{max}^E| \approx 2 \times 10^{-6}$ m$^3$ mol$^{-1}$ and $|K_{S,max}^E| \approx 12 \times 10^{-15}$ m$^3$ mol$^{-1}$ Pa$^{-1}$ at $T$ = 293.15 K for the mixture with *N*-methylpiperidine [9], while $|V_{max}^E| \approx 1.6 \times 10^{-6}$ m$^3$ mol$^{-1}$ and $|K_{S,max}^E| \approx 9 \times 10^{-15}$ m$^3$ mol$^{-1}$ Pa$^{-1}$ for that with piperidine. Similar correlations occur for aqueous solutions of pyridine and its methyl derivatives. The higher is the MP2 stabilization energy of the water-amine complex, the stronger is the propensity to the phase splitting [35] and the bigger are the negative excesses of volume [38] and of compression [43]. However, such regularities do not apply to the DFT/B3LYP and B2PLYP energies. This trend clearly shows, that the phase splitting of these aqueous systems is not due to the strong attraction between the like molecules, as it is assumed in the Bragg-Williams theory of regular solutions. The latter predicts phase splitting in terms of a lattice model, accounting only pairwise interaction energies. Thus, the origin of the concentration fluctuations and phase splitting in aqueous solutions of cyclic amines should be searched in the specific interactions between the cross-associates (or complexes). The aggregates of complexes may spontaneously expand due to cooperativity of the hydrogen bonds.

If the piperidine concentration is sufficiently small, then concentration fluctuations occur due to random changes of the aggregation numbers $n$ of particular complexes (C$_5$H$_{10}$N...HOH)$_n$, while the average number $\bar{n}$ remains constant at given $p$, $T$ and $x_1$. That would account for the Ornstein-Zernike scattering of the mixture with the water to piperidine molar ratio close to 25:1 (Figure 1, $x_1$ = 0.041). A maximum that is evident on the scattering curves at higher concentrations, $x_1 \geq 0.088$ (Figure 1), shows that the microheterogeneities gradually change from the statistical concentration fluctuations to a more structured form of a periodic ordering in the liquid with repeat distance related to the peak position [21]. Such behavior in binary mixtures is characterized as molecular microemulsions [44, 45], in analogy to the normal (water in oil) microemulsions for which the Teubner-Strey structure factor has been first applied [21]. Because of the shortage of water, bicontinuous structures may arise rather than typical "droplets" in the dispersion medium. In such case, water no longer fully surrounds the clusters of the 1:1 complexes, but remains rather in the neighborhood of the nitrogen atoms. This periodic order in aqueous solutions of piperidine makes them different from those of *N*-methylpiperidine, pyridine, 2-methylpyridine, 3-methylpiridine and 2,6-dimethylpyridine that exhibit only the concentration fluctuations [9, 38-40].

The above discussion of scattering curves concerned the mixtures with the piperidine mole fraction $x_1$ above 0.04, *i.e.* those containing 20 % and more of the amine by volume. Dilute aqueous solutions show several features that make them significantly different from the concentrated ones. Commonly for aqueous solutions of non-electrolytes and weak electrolytes, the isotherms of compressibility cross one another in a narrow range of the mole fractions $x_1$ from 0.017 to 0.027 (Figure 2). This may be interpreted in terms of the clathrate-like structure model, where the solvent-separated hydrophobic hydration makes the mixture compressibility approximately independent of temperature [6-8]. The piperidine guest molecules would be located in polyhedral cages formed by hydrogen-bonded water network. The polyhedra may either float freely in the surrounding bulk water [6] or form larger structures resembling the solid clathrate lattice [7, 8]. In the type II solid clathrate hydrate of piperidine + methane, the amine molecules occupy large hexakaidecahedral cages of the host water network, while methane is located in the smaller, dodecahedral ones [1]. The mole fraction of piperidine in water, calculated for that clathrate with larger voids filled with the amine molecules, is 0.056. That value delimits the maximum concentration above which the quantity of water is too small to form closed hydration shells around onefold piperidine molecules. Since the structure of liquid water is labile in comparison with that of solid ice or clathrate, the polyhedral hydration shells around piperidine molecules may exist in liquid solutions, even though the formation of the solid clathrate hydrates of piperidine requires a help gas under pressure [1]. Since the O–H...N bonds are relatively strong in comparison with the competing O–H...O ones, the nitrogen atom of the piperidine molecule would be hydrogen-bonded with the water shell.

The hydrophobic hydration accounts for the rapid decrease of the partial molar functions (*i.e.* volume and compression) of piperidine when the mole fraction $x_1$ approaches zero (Figure 5). At $x_1 < 0.1$ the dependence of $V_1$ and $K_{S,1}$ on $x_1$ is very steep, that results from the filling of the voids in water structure by the piperidine molecules. In this way, the polyhedra of water are stabilized and stiffened by the guest molecules. That results in the very low values of the limiting partial compression of piperidine, $\lim_{x_1 \to 0} K_{S,1} \approx 0$. Similar effect was observed for aqueous *N*-methylpiperidine [9], pyridine, 2-methylpiridine and 2,6-dimethylpyridine [43]. Contrary to the partial molar volume of *N*-methylpiperidine, which dependence on concentration shows distinct minimum at $x_1 \approx 0.01$ and $T = 293.15$ K, the $V_1$ of piperidine monotonically increases with $x_1$. Such minima of $V_1(x_1)$ are commonly attributed to the solvent-separated hydrophobic hydration [46]. The lack of minimum results probably from relative high hydrophilicity of piperidine, that is manifested in its dissociation constant higher than that of N-menthylpiperidine; the p*K*s of the two amines are 2.88 and 3.93, respectively [25]. When the amine concentration approaches zero, the dissociation degree increases. That adds significant "ionic" contribution to the volume contraction. The dissociation mechanism is probably quite complex, as it results from the studies of the Debye-type ultrasonic relaxation processes in dilute aqueous solutions of piperidine at $x_1 < 0.04$. A proton-transfer model, applied to account for the nanosecond-order-long relaxation times, required an assumption that piperidine molecule was hydrogen-bonded to two or three water molecules arranged in a line: $C_5H_{10}NH_2^+(H_2O)_{1\text{ or }2}OH^-$ [47].

Characteristic features of the partial volume and compression isotherms (Figure 5) are the concavities at $x_1 \approx 0.15$. Such concavity, or at least approximately linear segment of the function, are typical of aqueous solutions, reported also for *N*-methylpiperidine [9] and for pyridine and its methyl derivatives [43]. For pyridines in methanol, the $V_1$ and $K_{S,1}$ functions are convex in the whole concentration range [43]. Thus, the segments of the $V_1(x_1)$ and $K_{S,1}(x_1)$ functions with different slopes reflect various hydration modes: from the hydrophobic hydration at low $x_1$, through the aggregation of the complexes, mainly of the hydrophilic nature, up to dissolution of the 1:1 complexes in the amine solvent at $x_1 > 0.5$. In the latter concentration range, the partial volume and compression of the amine remains approximately constant and the system may be considered as a binary mixture of the pure amine and its 1:1 hydrate.

## 6. Conclusions

Molecular arrangement in piperidine – water mixtures is to some extent similar to that in *N*-methylpiperidine – water ones, suggested earlier [9]. In dilute aqueous solutions (mole fraction of the amine below 0.03), a solvent-separated hydration of piperidine occurs. Structures resembling semiclathrates arise, with the host amine molecules hydrogen-bonded to the water network. Direct proof of such structures in liquid state is not possible with the available experimental methods, but the presented thermodynamic results can be explained invoking their presence. With increasing concentration of piperidine, the clathrate-like polyhedra gradually decay, but the complexes of piperidine with water, $C_5H_{10}NH \cdot H_2O$, are still linked one to another through the O–H...O bonds between the hydration water molecules. In the piperidine-rich mixtures, the 1:1 complexes are dispersed uniformly in the amine.

A striking difference between the piperidine and *N*-methylpiperidine aqueous systems of higher concentrations is the periodic order of molecules in the first one rather than just the concentration fluctuations. That order is perhaps of the bicontinuous microemulsion type and it occurs at the broad range of molecular ratio of water to piperidine from 10:1 to 1:1. That is in spite of the lower MP2 stabilization energy of the 1:1 piperidine-water complexes in comparison with that of *N*-methylpiperidine ones [9]. The lower stabilization energy probably causes that piperidine is totally miscible with water, while the stronger water-amine interactions lead to phase splitting of the *N*-methylpiperidine – water system. Similar correlation between the cross-association energy and the propensity to phase separation of the binary mixtures with water was reported for pyridine, 2-methylpyridine and 2,6-dimethylpyridine [38].

**Acknowledgment**

This work has been supported by the European Commission under the 7th Framework Programme through the Key Action: Integrated Infrastructure Initiative for Neutron Scattering and Muon Spectroscopy: NMI3-II/FP7 – Contract No. 283883, and by the Russian Government Program of Competitive Growth of Kazan Federal University.

The GAUSSIAN09 calculations were carried out in the Academic Computer Centre CYFRONET of the University of Science and Technology in Cracow, ACC CYFRONET AGH, Kraków, Poland, http://www.cyfronet.pl, under Grant No. MNiSW/SGI3700/UŚląski/111/2007 and MNiSW/IBM_BC_HS21/ UŚląski /111/2007.

The GCMS analysis was performed by Dr. Andrzej Swinarew in the Institute of Materials Science, University of Silesia, Katowice, Poland, that is herein gratefully acknowledged.

# References

[1] W. Shin, S. Park, H. Ro, D.-Y. Koh, J. Seol, H. Lee, H, J. Chem. Thermodyn. 44 (2012) 20.

[2] W. Shin, S. Park, D.-Y. Koh, J. Seol, H. Ro, H. Lee, J. Phys. Chem. C 115 (2011) 18885.

[3] R. Susilo, S. Alavi, J. Ripmeester, P. Englezos, Fluid Phase Equilib. 263 (2008) 6.

[4] T. A. Strobel, C. A. Koh, E. D. Sloan, J. Phys. Chem. B 112 (2008) 1885.

[5] A. Martin, C. J. Peters, J. Phys. Chem. B 113 (2009) 7558.

[6] H. Endo, Bull. Chem. Soc. Jpn. 46 (1973) 1586.

[7] S. Ernst, J. Gliński, Materials Science III/3 (1977) 69.

[8] S. Ernst, W. Marczak, Bull. Pol. Acad. Sci. Chem. 43 (1995) 259.

[9] W. Marczak, M. Łężniak, M. Zorębski, P. Lodowski, A. Przybyła, D. Truszkowska, L. Almásy, RSC Advances 3 (2013) 22053.

[10] Y. Coulier, K. Ballerat-Busserolles, L. Rodier, J.-Y. Coxam, Fluid Phase Equilib. 296 (2010) 206.

[11] R. M. Stephenson, J. Chem. Eng. Data 38 (1993) 428.

[12] E. Álvarez, D. Gómez-Díaz, D. La Rubia, J. M. Navaza, J. Chem. Eng. Data 50 (2005) 1829.

[13] W. Afzal, A. Valtz, Ch. Coquelet, D. Richon, J. Chem. Thermodyn. 40 (2008) 47.

[14] D. Gómez-Díaz, J. M. Navaza, J. Chem. Eng. Data 51 (2006) 722.

[15] NIST Chemistry WebBook, NIST Standard Reference Database Number 69 http://webbook.nist.gov/chemistry/ (retrieved 05.02.2014)

[16] T. E. Daubert, R. P. Danner, H. M. Sibel, C. C. Stebbins, Physical and Thermodynamic Properties of Pure Chemicals. Data Compilation, Taylor & Francis, Washington, DC, 1997. (cited in [13])

[17] A. Brûlet, D. Lairez, A. Lapp, J.-P. Cotton, J. Appl. Crystallogr. 40 (2007) 165.

[18] W. Marczak, K. Kiełek, B. Czech, H. Flakus, M. Rogalski, Phys. Chem. Chem. Phys. 11 (2009) 2668.

[19] W. Marczak, J. Acoust. Soc. Am. 102 (1997) 2776.

[20] S. Ernst, W. Marczak, D. Kmiotek, J. Chem. Eng. Data 41 (1996) 128.

[21] M. Teubner, R. Strey, J. Chem. Phys. 87 (1987) 3195.

[22] G. D'Arrigo, R. Giordano, J. Teixeira, Eur. Phys. J. E 29 (2009) 37.

[23] J. Kohlbrecher, SASfit software package, available at http://kur.web.psi.ch/sans1/SANSSoft/sasfit.html

[24] A. Das, M. Frenkel, N. A. M. Gadalla, S. Kudchadker, K. N. Marsh, A. S. Rodgers, R. C. Wilhoit, J. Phys. Chem. Ref. Data 22 (1993) 659.

[25] H. Całus, Termodynamika Chemiczna. Termochemia (Part B, Chap. 3) in: A. Dorabialska, Z. Łada, M. Michalski, W. Rodewald (co-ordinators), Poradnik Fizykochemiczny, Wydawnictwa Naukowo-Techniczne, Warszawa 1974.

[26] R. A. Kendall, T. H. Dunning, R. J. Harrison, J. Chem. Phys. 96 (1992) 6796.

[27] T. H. Dunning, J. Chem. Phys. 90 (1989) 1007.

[28] A. D. Becke, J. Chem. Phys. 98 (1993) 5648.

[29] C. Lee, W. Yang, R. G. Parr, Phys. Rev. B: Condens. Matter Mater. Phys. 37 (1988) 785.

[30] S. Grimme, J. Chem. Phys. 124 (2006) 034108.

[31] M. J. Frisch et al., GAUSSIAN 09, Revision A.02, GAUSSIAN, Inc., Wallingford CT, 2009.

[32] A. Przybyła, P. Kubica, Sz. Bacior, P. Lodowski, W. Marczak, Chem. Phys. Lett. 512 (2011) 199.

[33] U. Spoerel, W. Stahl, Chem. Phys. 239 (1998) 97.

[34] P. R. Rablen, J. W. Lockman, W. L. Jorgensen, J. Phys. Chem. A 102 (1998) 3782.

[35] I. Pápai, G. Jancsó, J. Phys. Chem. A 104 (2000) 2132.

[36] J. Smets, W. McCarthy, G. Maes, L. Adamowicz, J. Mol. Struct., 476 (1999) 27.

[37] V. Dohnal, A. H. Roux, V. Hynek, J. Solution Chem. 23 (1994) 889.

[38] W. Marczak, B. Czech, L. Almásy, D. Lairez, Phys. Chem. Chem. Phys. 13 (2011) 6260.

[39] L. Almásy, L. Cser, G. Jancsó, J. Mol. Liquids 101 (2002) 89.

[40] L. Almásy, G. Jancsó, J. Mol. Liquids 113 (2004) 61.

[41] K Nishikawa, Y Kasahara, T Ichioka, J. Phys. Chem. B 106, (2002) 693.

[42] L. Almásy, M. Turmine, A. Perera, J Phys. Chem. B 112, (2008) 2382.

[43] B. Czech, P. Lodowski, W. Marczak, Chem. Phys. Lett. 556 (2013) 132.

[44] B. Kezic, A. Perera, J. Chem. Phys. 137, (2012) 014501.

[45] B. Kezic, A. Perera, J. Chem. Phys. 137, (2012) 134502.

[46] F. Franks, Water: 2$^{nd}$ edition. A matrix of life, The Royal Society of Chemistry, Cambridge 2000.

[47] S. Nishikawa, N. Arakane, N. Kuramoto, J. Phys. Chem. 99 (1995) 369.

Table 1

The quasi-periodic repeat distance $D$, Eq. (2), and the decay length of the correlations $\xi$, Eq. (3), in the mixtures of piperidine with $D_2O$

| $x_1$ | $D$ / nm | $\xi$ / nm |
|---|---|---|
| 0.044 | 3.95 ± 0.21 | 0.31 ± 0.01 |
| 0.088 | 2.58 ± 0.01 | 0.53 ± 0.01 |
| 0.106 | 2.36 ± 0.01 | 0.59 ± 0.01 |
| 0.166 | 1.99 ± 0.01 | 0.79 ± 0.01 |
| 0.391 | 1.36 ± 0.02 | 0.55 ± 0.05 |
| 0.474 | 1.28 ± 0.04 | 0.57 ± 0.09 |
| 0.588 | 0.92 ± 0.75 | 0.37 ± 0.45 |

Table 2

Association energies and selected structural parameters[a] of the 1:1 complex of piperidine with water and of the piperidine dimer

|  | DFT/B3LYP | | | MP2[b] | | | B2PLYP | | |
|---|---|---|---|---|---|---|---|---|---|
|  | $\Delta E$ | $\Delta E_{BSSE}$ | $\Delta E_{BSSE+ZPE}$ | $\Delta E$ | $\Delta E_{BSSE}$ | $\Delta E_{BSSE+ZPE}$ | $\Delta E$ | $\Delta E_{BSSE}$ | $\Delta E_{BSSE+ZPE}$ |
| HO–H...NHC$_5$H$_{10}$ | 30.1 | 27.1 | 17.5 | 39.7 | 31.5 | 21.9 | 34.0 | 29.4 | 20.3 |
| C$_5$H$_{10}$N–H...N(H)C$_5$H$_{10}$ | 12.5 | 9.1 | 6.4 | 29.1 | 19.2 | 16.4 | 19.6 | 13.0 | 10.8 |
|  | $R_{NAH}$ | $\alpha_{C(4)NH}$ | $\alpha_{XHN}$[c] |  |  |  | $R_{NAH}$ | $\alpha_{C(4)NH}$ | $\alpha_{XHN}$[c] |
| HO–H...NHC$_5$H$_{10}$ | 1.909 | 97.5 | 170.9 |  |  |  | 1.900 | 95.2 | 169.4 |
| C$_5$H$_{10}$N–H...N(H)C$_5$H$_{10}$ | 2.275 | 109.7 | 171.8 |  |  |  | 2.211 | 101.5 | 179.1 |

BSSE – corrected for the basis set superposition error, ZPE – corrected for the zero-point vibrational energy

[a] Energies $\Delta E$ in kJ mol$^{-1}$, bond lengths $R$ in Å, valence angles $\alpha$ in degrees

[b] Results obtained for single point calculations from DFT/B3LYP optimized geometries

[c] X = O for H$_2$O and X = N for C$_5$H$_{10}$NH as the donors of proton

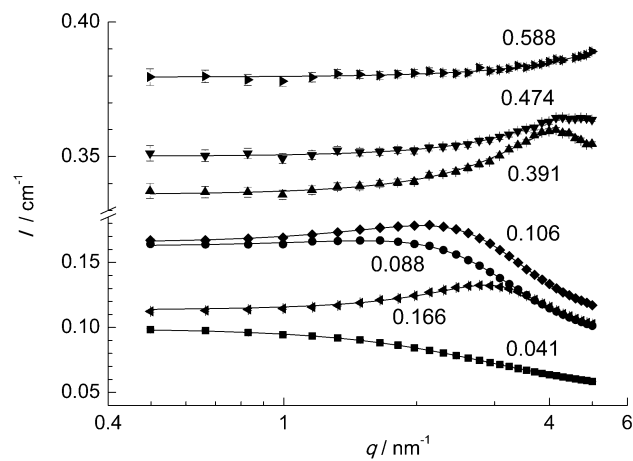

**Figure 1.** SANS in the mixtures of piperidine (1) with $D_2O$ (2) at $T = 298.15$ K. Points – experimental results, lines – Teubner-Strey Eq. (1) with coefficients reported in Table 1. The mole fractions of the binary mixtures are given in the line labels.

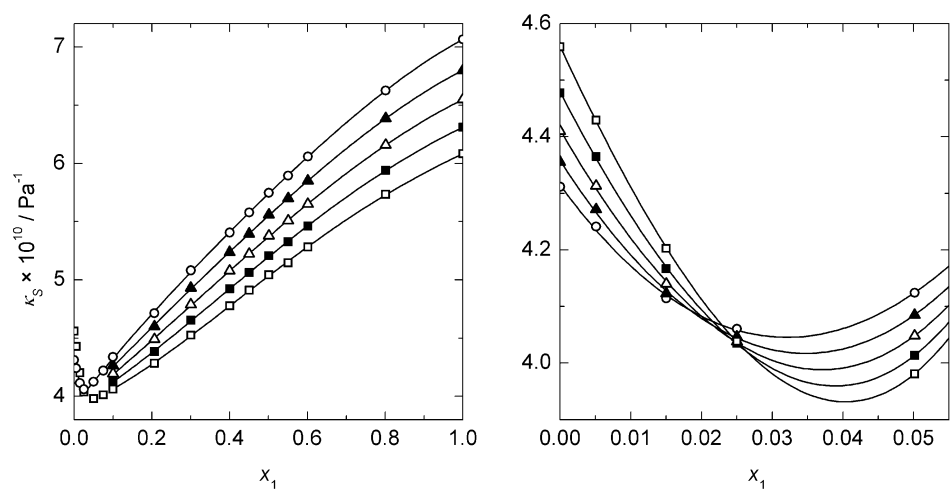

**Figure 2.** Isentropic compressibility of binary mixtures of piperidine (1) with water (2). Points – experimental results: □ – $T = 293.15$ K, ■ – $T = 298.15$ K, Δ – $T = 303.15$ K, ▲ – $T = 308.15$ K, ○ – $T = 313.15$ K; lines – interpolation. Left – whole range of the mole fractions (some points were omitted for the picture clarity); right – dilute aqueous solutions.

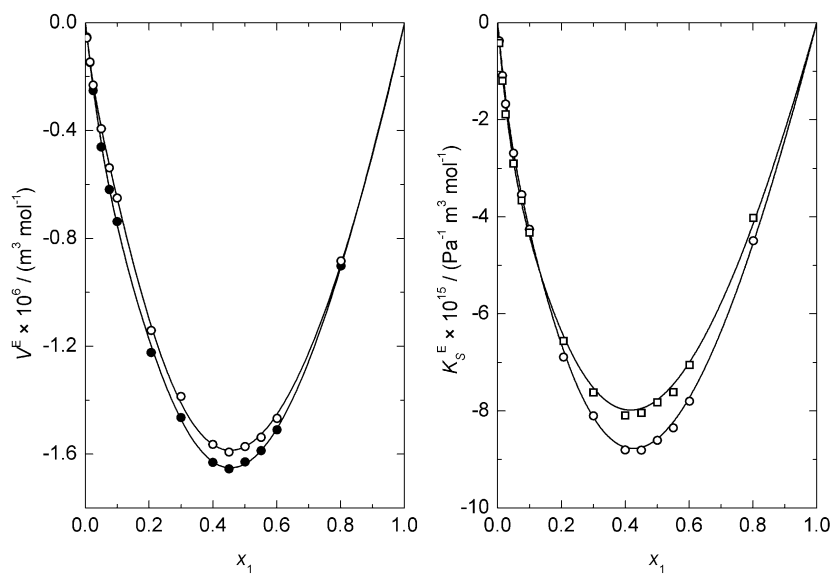

**Figure 3.** Excesses of the molar volume, $V^E$, and of the molar isentropic compression, $K_S^E$, of piperidine (1) + water (2). Points – experimental results: ● – $T$ = 288.15 K, □ – $T$ = 293.15 K, ○ – $T$ = 313.15 K; lines – Eq. (5).

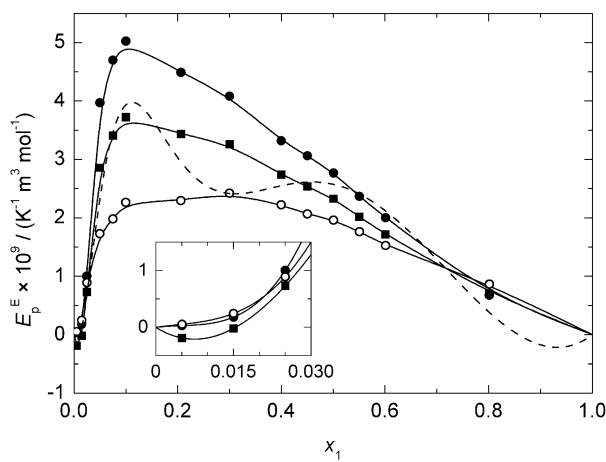

**Figure 4.** Excess molar thermal expansion of piperidine (1) + water (2). Points – experimental results: ● – $T$ = 288.15 K, ■ – $T$ = 298.15 K, ○ – $T$ = 313.15 K; solid lines – cubic spline interpolation, broken line – Eq. (6) with $V^E$ given by Eq. (5).

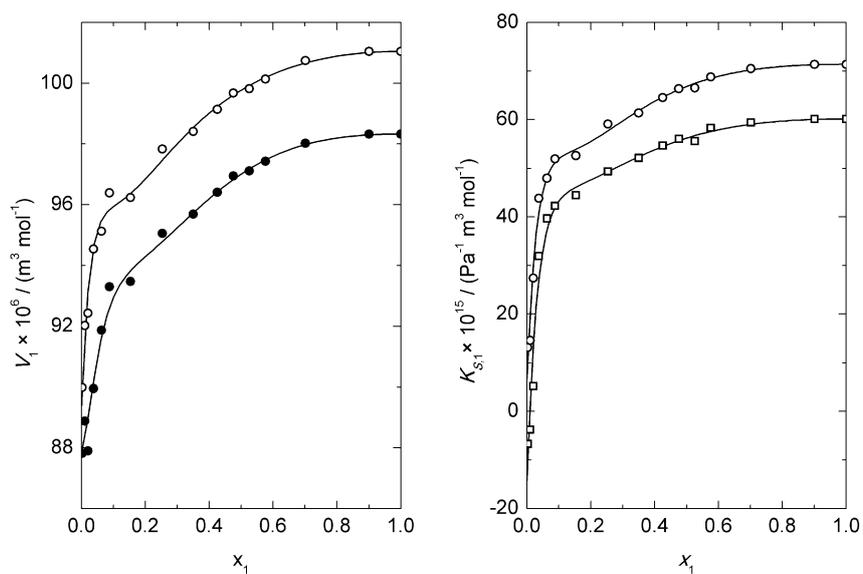

**Figure 5.** Partial molar functions of piperidine (1) in binary mixtures with water (2): volume $V_1$ and isentropic compression $K_{S,1}$. Points – Eq. (8): ● – $T$ = 288.15 K, □ – $T$ = 293.15 K, ○ – $T$ = 313.15 K; lines – Eq. (7).

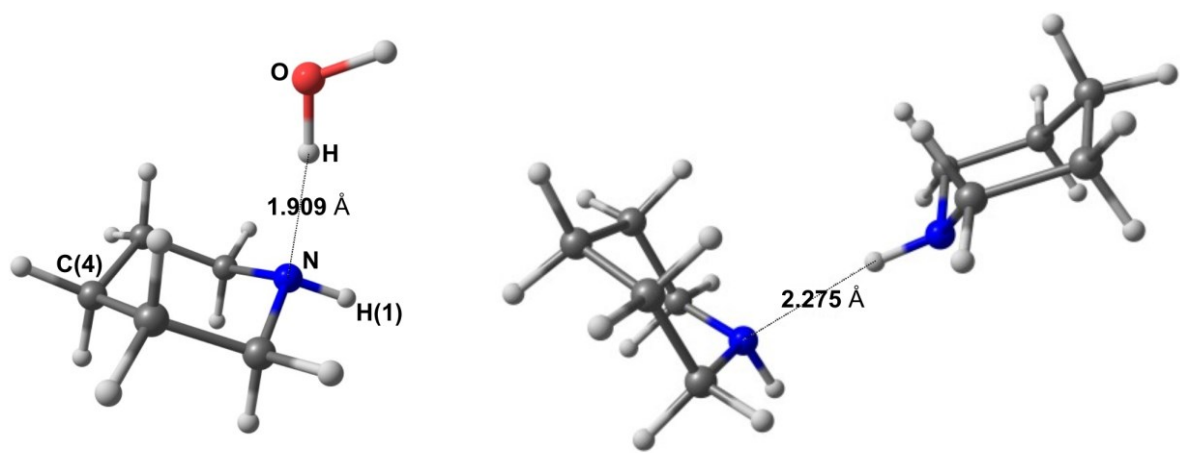

**Figure 6.** Geometry of the 1:1 complex of piperidine with water and of the piperidine dimer calculated by the B3LYP/aug-cc-pVDZ method. Bond lengths and valence angles reported in Table 2.

19